\documentclass[twocolumn]{aastex631}
\usepackage{rotating}
\usepackage{booktabs}
\usepackage{amsmath,amsthm,amsfonts,amssymb,extarrows}
\usepackage{apjfonts}
\usepackage{float}
\usepackage{xcolor}
\usepackage{xspace}

\definecolor{mBlue}{RGB}{51, 77, 167}

\hypersetup{
	colorlinks,
	citecolor=blue,
	linkcolor=magenta,
	urlcolor=blue,
}

\newcommand{\chandra}{\textit{Chandra}\xspace}
\newcommand{\xmm}{\textit{XMM-Newton}\xspace}

\received{2024 January 11; revised 2024 April 2; accepted 2024 April 28; published 2024 June 28}

\shorttitle{\ion{Fe}{17} Wavelengths}
\shortauthors{Chintan~Shah~{et~al.}}

\begin{document}

\title{\Large{{High-Precision Transition Energy Measurements of Neon-like \ion{Fe}{17} Ions}}}

\correspondingauthor{Chintan Shah}
\email{chintan.shah@mpi-hd.mpg.de}

\author[0000-0002-6484-3803]{Chintan~Shah}
\affiliation{NASA Goddard Space Flight Center, 8800 Greenbelt Rd, Greenbelt, MD 20771, USA}
\affiliation{Max-Planck-Institut f\"ur Kernphysik, Saupfercheckweg 1, 69117 Heidelberg, Germany}
\affiliation{Center for Space Sciences and Technology, University of Maryland, Baltimore County, 1000 Hilltop Circle, Baltimore, MD 21250, USA }

\author[0000-0003-4571-2282]{Moto~Togawa}
\affiliation{Max-Planck-Institut f\"ur Kernphysik, Saupfercheckweg 1, 69117 Heidelberg, Germany}
\affiliation{European XFEL, Holzkoppel 4, 22869 Schenefeld, Germany}
\affiliation{Heidelberg Graduate School of Fundamental Physics, Ruprecht-Karls-Universität Heidelberg, Im Neuenheimer Feld 226, 69120 Heidelberg, Germany}

\author[0009-0003-8574-4542]{Marc~Botz}
\affiliation{Max-Planck-Institut f\"ur Kernphysik, Saupfercheckweg 1, 69117 Heidelberg, Germany}
\affiliation{Heidelberg Graduate School of Fundamental Physics, Ruprecht-Karls-Universität Heidelberg, Im Neuenheimer Feld 226, 69120 Heidelberg, Germany}%

\author[0009-0002-2934-6016]{Jonas~Danisch}
\affiliation{Max-Planck-Institut f\"ur Kernphysik, Saupfercheckweg 1, 69117 Heidelberg, Germany}

\author[0009-0006-2461-9571]{Joschka~J.~Goes}
\affiliation{Max-Planck-Institut f\"ur Kernphysik, Saupfercheckweg 1, 69117 Heidelberg, Germany}

\author[0000-0002-1976-5121]{Sonja~Bernitt}
\affiliation{GSI Helmholtzzentrum für Schwerionenforschung, Planckstraße 1, 64291 Darmstadt, Germany}%
\affiliation{Helmholtz-Institut Jena, Fröbelstieg 3, 07743 Jena, Germany}
\affiliation{Institut für Optik und Quantenelektronik, Friedrich-Schiller-Universität, Max-Wien-Platz 1, 07743 Jena, Germany}
\affiliation{Max-Planck-Institut f\"ur Kernphysik, Saupfercheckweg 1, 69117 Heidelberg, Germany}

\author[0009-0005-9076-3101]{Marleen~Maxton}
\affiliation{Max-Planck-Institut f\"ur Kernphysik, Saupfercheckweg 1, 69117 Heidelberg, Germany}

\author{Kai~K\"{o}bnick}
\affiliation{Max-Planck-Institut f\"ur Kernphysik, Saupfercheckweg 1, 69117 Heidelberg, Germany}

\author{Jens~Buck}
\affiliation{Institut f\"ur Experimentelle und Angewandte Physik, Christian-Albrechts-Universität zu Kiel, Kiel, Germany}

\author[0009-0009-7545-2101]{J\"orn~Seltmann}
\affiliation{Deutsches Elektronen-Synchrotron (DESY), Notkestrasse 85, 22607 Hamburg, Germany}

\author[0000-0002-0114-2110]{Moritz~Hoesch}
\affiliation{Deutsches Elektronen-Synchrotron (DESY), Notkestrasse 85, 22607 Hamburg, Germany}

\author[0000-0001-9136-8449]{Ming~Feng~Gu}
\affiliation{Space Science Laboratory, University of California, Berkeley, CA 94720, USA}%

\author[0000-0002-6374-1119]{F.~Scott~Porter}
\affiliation{NASA Goddard Space Flight Center, 8800 Greenbelt Rd, Greenbelt, MD 20771, USA}%

\author[0000-0002-5312-3747]{Thomas~Pfeifer}
\affiliation{Max-Planck-Institut f\"ur Kernphysik, Saupfercheckweg 1, 69117 Heidelberg, Germany}%
   
\author[0000-0002-3331-7595]{Maurice~A.~Leutenegger}
\affiliation{NASA Goddard Space Flight Center, 8800 Greenbelt Rd, Greenbelt, MD 20771, USA}%

\author[0000-0002-3724-3730]{Charles~Cheung}
\affiliation{Department of Physics and Astronomy, University of Delaware, Newark, Delaware 19716, USA}

\author[0000-0002-1305-4011]{Marianna~S.~Safronova}
\affiliation{Department of Physics and Astronomy, University of Delaware, Newark, Delaware 19716, USA}
 
\author[0000-0002-2937-8037]{Jos\'e~R.~{Crespo~L\'opez-Urrutia}}
\affiliation{Max-Planck-Institut f\"ur Kernphysik, Saupfercheckweg 1, 69117 Heidelberg, Germany}

\begin{abstract}
We improve by a factor of 4--20 the energy accuracy of the strongest soft X-ray transitions of \ion{Fe}{17} ions by resonantly exciting them in an electron beam ion trap with a monochromatic beam at the P04 beamline of the PETRA III synchrotron facility. By simultaneously tracking instantaneous photon-energy fluctuations with a high-resolution photoelectron spectrometer, we minimize systematic uncertainties down to 10--15 meV, or velocity equivalent $\pm\sim$5 km\,s$^{-1}$ in their rest energies, substantially improving our knowledge of this key astrophysical ion. Our large-scale configuration-interaction computations include more than four million relativistic configurations and agree with the experiment at a level without precedent for a 10-electron system. Thereby, theoretical uncertainties for interelectronic correlations become far smaller than those of quantum electrodynamics (QED) corrections. The present QED benchmark strengthens our trust in future calculations of many other complex atomic ions of interest to astrophysics, plasma physics, and for the development of optical clocks with highly charged ions. 
\end{abstract}

\section{Introduction}\label{sec:intro}

Over the past three decades, extensive research has focused on the soft X-ray emission from Ne-like iron (\ion{Fe}{17}, Fe$^{16+}$), particularly in hot astrophysical plasmas observed by \chandra and \xmm~\citep{behar2001chandra,brinkman2001first}. The dominant spectral transitions $3d\rightarrow2p$ and $3s\rightarrow2p$ of \ion{Fe}{17} within the 700--850 eV range (14.5 to 17.5 \AA) play a crucial role in deducing the plasma parameters across various sources. These parameters include the electron temperature, density, elemental abundance, gas motion, and photon scattering opacity~\citep{parkinson1973new, smith1985,schmelz1992,waljeski1994,phillips1996,behar2001chandra,mauche2001,doron2002,xpb2002,gu2003FeL,pfk2003,werner2009,pradhan2011atomic,beiersdorfer2018,gu2019,gu2020capellaEBIT,grell2021}.

Despite decades of study, since early solar X-ray observations~\citep{parkinson1973new,smith1985,schmelz1992,waljeski1994}, discrepancies between observed and theoretical intensity ratios~\citep{brown1998laboratory} have persisted. Early explanations invoking resonance scattering~\citep{mckenzie1980solar,schmelz1992,saba1999} found no confirmation in measurements with electron-beam ion traps (EBITs) and tokamaks that also agreed with solar observations~\citep{brown1998laboratory,brown2001diagnostic,brown2001systematic,brown2006energy,beiersdorfer2002laboratory,gillaspy2011fe,beiersdorfer2004laboratory,beiersdorfer2017,shah2019}. As optically thin laboratory plasmas are not subject to resonance scattering, indirect line formation mechanisms were suggested~\citep{cpr2002,gu2003FeL,beiersdorfer2008,beiersdorfer2014,beiersdorfer2015,shah2019,grilo2021,gu2020capellaEBIT}. An experiment with a free-electron laser aimed at directly determining the oscillator-strength ratio for lines 3C and 3D without uncertainties due to electron-impact excitation. Its unexpected results departing even more from theory were attributed to inaccuracies in calculated oscillator strengths~\citep{bernitt2012unexpectedly}, but soon after, unforeseen transient nonequilibrium effects and population transfer due to the ultrabrilliant peak photon flux explained them~\citep{ock2014,oreshkina2016x,Loch_2015,wu2019change}. Our later measurements~\citep{kuhn2020high} with synchrotron radiation avoided this nonlinear systematic and improved the accuracy of the oscillator-strength ratio while still disagreeing with the theory. Finally, further increases in resolving power and signal-to-noise ratio found the cause of the persistent discrepancies in hitherto unresolvable line wings and diffraction effects and brought the oscillator-strength ratio in line with state-of-the-art predictions~\citep{kuhn2022}.

In spite of these advances, many questions remain open for this essential ion and many other less-studied species. For instance, widely used wavelength references for \ion{Fe}{17} from EBIT measurements using a crystal spectrometer with a resolving power of 500--700 have uncertainties of 1--3~m\AA\ ($\sim$ 40--200 meV), Doppler-equivalent to $\sim$15--50~km\,s$^{-1}$, i.e. $\sim$60--180 parts per million (ppm), for $n=3-2$, and double that for high-$n$ transitions \citep{wargelin1994, brown1998laboratory}. This is only marginally adequate for analysis of high-resolution diffraction grating spectra acquired with the \chandra High Energy Transmission Grating Spectrometer (HETGS), which can measure velocities of bright emission lines with $\sim$25~km\,s$^{-1}$ systematic uncertainty \citep{Ishibashi2006, Bozzo2023}. These uncertainties in transition energies will also impair the achievement of the science goals of other extant, upcoming, and proposed missions, including \xmm~\citep{jansen2001xmm,den2001reflection},~\textit{XRISM}~\citep{xrism2018}), \textit{Athena}~\citep{pajot2018athena,barret2016},~\textit{LEM}~\citep{lem2022}, \textit{Arcus}~\citep{arcus_instrum,arcus_science},~and~\textit{Lynx}~\citep{Schwartz2019}. Even though some of these missions feature spectrometers with full-width-at-half-maximum (FWHM) resolution far broader than the uncertainties of published rest-energy determinations, well-exposed spectra of bright objects with a high signal-to-noise ratio will allow centroid determination with uncertainties comparable to or smaller than these prior measurements. There is clearly a need for better determinations of the \ion{Fe}{17} transition energies that will allow us to take full advantage of the resolving power of current and future missions, as well as improved and well-benchmarked theoretical methods that can provide energies for transitions that have not yet been measured with sufficient precision.

We report new measurements of the rest energies of key \ion{Fe}{17} transitions with an EBIT at the P04 beamline of the PETRA III synchrotron with uncertainties below 15 ppm, an improvement by a factor of 4--20 over the status quo. The accuracy of our results translates in velocity terms to 5~km\,s$^{-1}$, and fully unlocks the value of archived and forthcoming observations from XMM-Newton and Chandra, as well as of accurate velocimetry targeted by upcoming missions~\citep{pajot2018athena,barret2016,lem2022,arcus_instrum,arcus_science}. We also test the large-scale configuration interaction (CI) approach and, therefore, our combination of the CI and coupled-cluster approaches (CI\,+\,all-order method), which is crucial for the development of high-precision clocks~\citep{HCIs}, and essential for understanding the quantum electrodynamics (QED) effects in many-electron systems. By applying the model potential approach~\citep{tupitsyn2016} using the QEDMOD package~\citep{shabaev2018}, we incorporate QED effects into the effective Hamiltonian, basis-set orbitals, and one-electron matrix elements---a widely employed practice. The quality of the QED model potential is usually assessed against exact solutions for H-like ions since the uncertainty in the electronic correlation in HCI with a few valence electrons is usually larger, or at the level of QED contributions unless the ionization degree is rather high. Until this work, there were no estimates regarding the accuracy of the QEDMOD approach for the majority of many-electron systems.

Motivated by our highly accurate experimental results, we carry out new CI computations, taking the contributions from high $nl$ states into account, increasing the number of relativistic configurations from 1.2 million in our previous work~\citep{kuhn2022,cheung2021scalable} to over 4 million, and investigating the convergence of the computations in both of these parameters. The results show a remarkable degree of numerical convergence across all energy levels and agree with the measurements to a level of 1--33 meV (1--45 ppm) that is unprecedented for a complex ion such as \ion{Fe}{17}. For the first time, uncertainties in the electronic correlations smaller than QED corrections allow us to test the accuracy of the QED contribution in a many-electron system.  

\begin{figure*}[ht]
\centering
\includegraphics[clip=true,width=\textwidth]{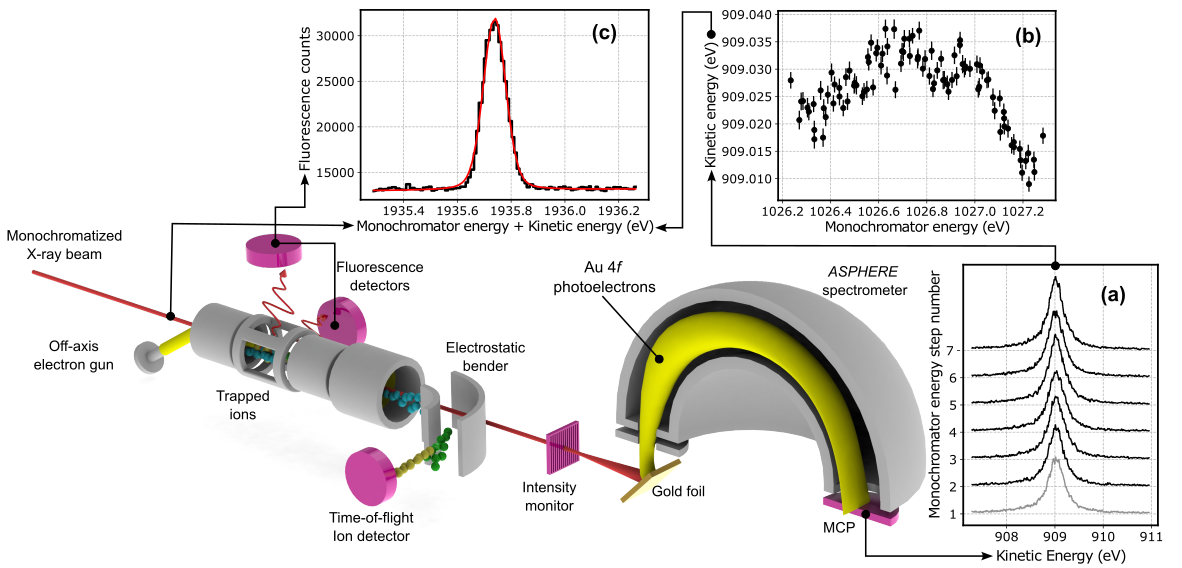} 
\caption{A photon beam of variable energy excites an elongated ion ensemble within a portable electron beam ion trap, PolarX-EBIT \citep{micke2018}. Emitted fluorescence X-rays are recorded by two silicon drift detectors. Ions periodically released from the trap are mass-analyzed by their time of flight as a monitoring diagnostic of the trapped ion content. Downstream, the photon beam passes through a wire mesh used to measure its intensity before hitting a gold target and releasing photoelectrons that enter ASPHERE, a high-resolution hemispherical electron-energy analyzer. ASPHERE records (a) the Voigt-like kinetic energy distribution of Au $4f_{7/2}$ electrons and their centroids (b) at each monochromator energy step. Because we apply a bias to the Au target that tracks changes in the nominal monochromator energy, ideally, (b) is expected to exhibit constant values, but it shows small yet reproducible periodic deviations from the nominal monochromator energy scale due to interpolation errors in the angular encoders. These deviations are corrected for in (c) prior to calibration with reference lines.}
\label{fig:exp}
\end{figure*}

\section{Measurements and Data Analysis}
\label{sec:exp}

PolarX-EBIT~\citep{micke2018} was designed for the study of highly charged ions interacting with X-ray photons at synchrotrons and free-electron lasers \citep[see][]{Leutenegger2020,Togawa2020,kuhn2020high,kuhn2022,steinbruegge2022, Stierhof2022}. Its off-axis electron gun emits a nearly monoenergetic electron beam that is compressed to a diameter of less than 100~\textmu m by a magnetic field of $\sim$870~mT generated by permanent magnets. Considering the overlap of the ion cloud and the electron beam leads to an effective electron density of $\sim10^{10}\,\mathrm{cm}^{-3}$. Iron pentacarbonyl (Fe(CO)$_5$) molecules enter the trap region as a tenuous beam through a two-stage differential pumping system. There, electron-impact dissociation generates Fe atoms, and step-wise electron-impact ionization produces highly charged ions that remain radially trapped by the ensuing negative space-charge potential of the electron beam, and axially by biased cylindrical drift tubes. We chose operating conditions to ensure that \ion{Fe}{17} ions mostly populate the trap.

At the soft X-ray beamline P04, an APPLE II undulator~\citep{viefhaus2013variable} produces circularly polarized photons, which are then sent through a monochromator equipped with a variable line-spacing grating of 1200 lines\,mm$^{-1}$ mean groove density. Using an exit slit opening of 50~\textmu m, the energy resolution $\Delta$E was set to a value of approximately E/$\Delta$E$\approx$13,000 in the energy range of 700--1100 eV. A pair of plane-elliptical mirrors refocus this beam onto the ion cloud. The photon beam energy is scanned over the \ion{Fe}{17} transitions of interest and the corresponding calibration lines. Two silicon drift detectors (SDD) mounted at the top and on the side of the EBIT register fluorescence, with $\sim$100 eV FWHM resolution, following from resonant photoexcitation as well as electron-impact excitation.

To calibrate the monochromator photon-energy scale, we excite K-shell transitions in H-like and He-like oxygen, fluorine, and neon ions trapped in PolarX-EBIT. Their energies can be calculated with uncertainties well below 1\,meV. We take values for the H-like $1s\rightarrow 2p$ transitions from \cite{yerokhin2015}, and from \cite{erickson1977} for $1s\rightarrow np$ up to $n=7$, and for He-like ions, we take energy values from \cite{yerokhin2019} for $1s-np$ transitions up to $n=7$. 

The monochromator disperses the spectrum of the undulator cone on the exit slit by choice of incidence and diffraction angles of the grating, which is accomplished by appropriate rotations of both the grating and mirror, with the extra degree of freedom removed by requiring fulfillment of the constant fix-focus (CFF) condition~ \citep{Follath2001}. The absolute angles of both grating and mirror are recorded using angular encoders. To measure electronic transitions with narrow natural line widths < 50\,meV, angular increments as small as $\approx10^{-5}$ degrees have to be resolved, equivalent to 36 milliarcseconds or 175 nrad. The installed Heidenhain RON 905 angular encoders have 36\,000 reference marks per turn, or one every 10$^{-2}$ degrees (36 arcseconds). Such encoders interpolate angle changes a thousand times between each mark. An LED source is positioned on one side of the disk, while two photodiodes are positioned on the opposite side of the disk to record the light intensity modulated by slight rotations of the encoder disk.\footnote{\url{https://www.heidenhain.us/wp-content/uploads/2022/07/591109-24_Angle_Encoders_with_Integral_Bearing.pdf}} These intensity variations are then stored in an empirical lookup table in the hardware. However, the process is highly sensitive to imperfections in the analog signals, which can lead to periodic subdivision errors~\citep{Krempasky2011,follath2010}. Furthermore, within each monochromator, there are two encoders--one dedicated to the grating and the other to the focusing mirror. This doubles the uncertainty in the interpolation, impacting the determination of the diffraction angle. Consequently, this can cause the nominal monochromator energy to deviate from the actual photon beam energy. This problem was previously observed in our studies at P04 \citep{kuhn2022,togawa2023} and other beamlines~\citep{follath2010,Krempasky2011}, and leads to periodic fluctuations in the nominal photon-energy scale, which in our case have peak-to-peak amplitudes of up to $\sim$50 meV below 900 eV and $\sim$70 meV above 900 eV.

\begin{figure*}
    \centering
    \includegraphics[clip=true,width=0.9277\textwidth]{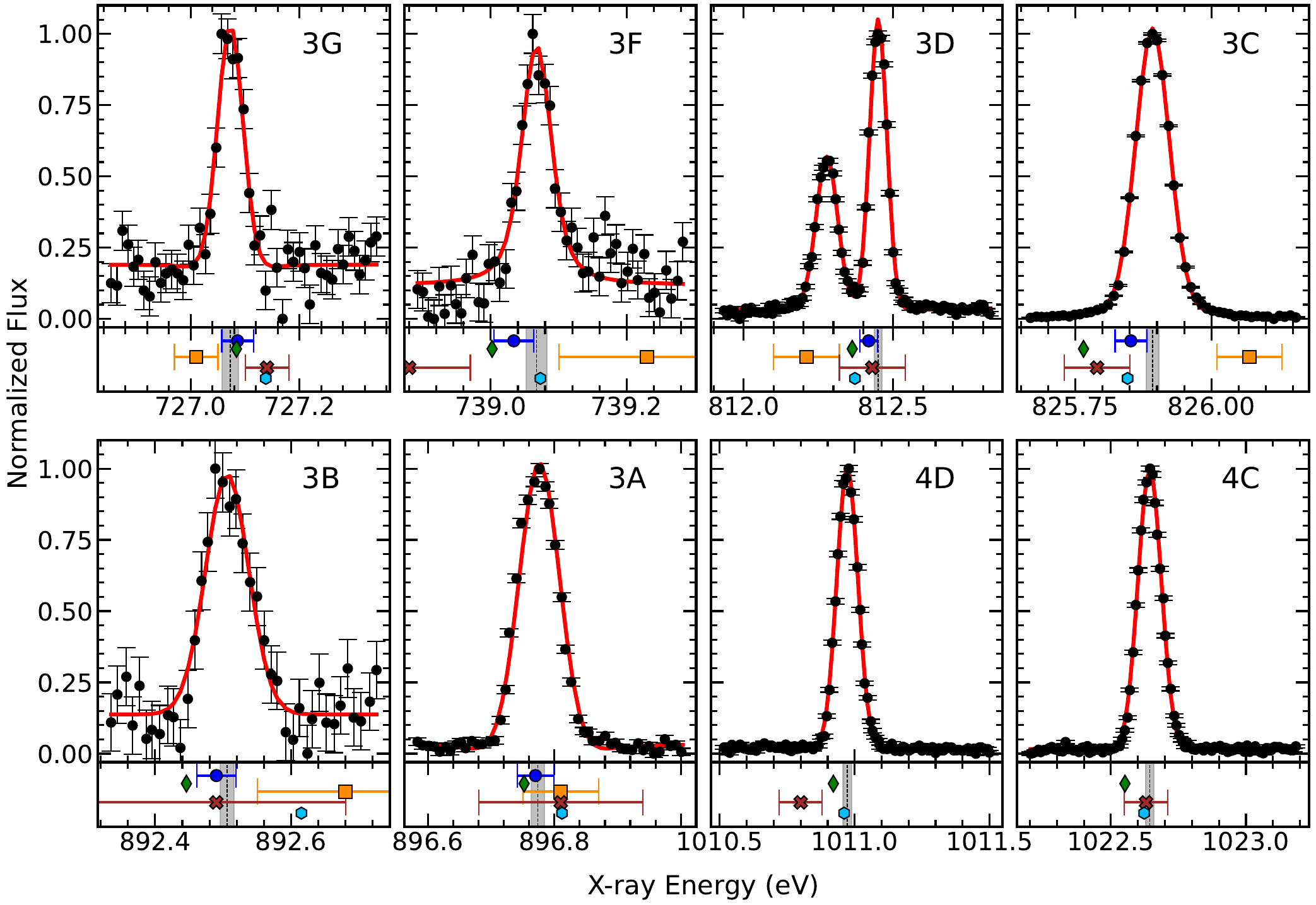} 
    \caption{\label{fig:sp}
    (Top panels) Representative scan for each of the measured Fe XVII lines. Both the data and model are scaled to the range [0,1]. 
    (Bottom panels) Measured transition energies derived from a weighted average of all scans are depicted with their total uncertainty represented by the gray band. They are compared against Large CI (blue circles), FAC MBPT (green diamonds), solar observations (cyan hexagones), and previous laboratory data (orange squares:~\cite{wargelin1994}, brown crosses:~\cite{brown1998laboratory}).}
\end{figure*}

To correct for them while scanning the monochromator to excite resonant transitions, we direct the photon beam exiting the EBIT onto a gold target mounted on a high-resolution hemispherical electron-energy analyzer, ASPHERE~\citep{ROSSNAGEL2001}, as shown in Fig.~\ref{fig:exp}. There, $4f_{5/2,7/2}$ photoelectrons are emitted, and their kinetic energy is measured. The kinetic energy of these photoelectrons is given by the difference between the photon energy and the binding energy of the electrons, along with any potential bias applied to the gold target. If the bias potential applied to the target is constant, any change in the photon energy will manifest itself as a change in the kinetic energy of the $4f$ Au electrons. However, if we change the target bias to track changes in the nominal photon energy, the electron kinetic energy remains nominally constant and the photoelectron peak (Au $4f_{7/2}$) can appear at a fixed position on the electron detector, see Fig.~\ref{fig:exp}\,(a). Thus, any deviation of the actual photon energy from the nominal photon energy set by the monochromator would result in a deviation of the kinetic energy of the photoelectrons. An example of such a deviation, reflecting the interpolation inaccuracies of the two angular encoders of the monochromator, is shown in Fig.~\ref{fig:exp}\,(b). We fit Au $4f_{7/2}$ peaks (line widths of about 700 meV FWHM) with Voigt profiles to find their centroids and determine their kinetic energies with uncertainties of a few meV at electron count rates of $\sim$10$^{4}/\mathrm{s}$. By cooling the gold target to liquid nitrogen temperature, we further reduce the peak width to $\sim$450 meV, which further improves centroid determination. We then use this information to correct each step of the nominal monochromator energy scale. To avoid any assumptions in modeling these deviations shown in Fig.~\ref{fig:exp}\,(b), and because the addition of an arbitrary constant term to the energy scale will be removed when calibrating against known reference energies, we simply add the measured electron kinetic energy directly to the nominal monochromator energy scale rather than first subtracting a nominal kinetic energy offset. On this corrected monochromator energy scale (Fig.~\ref{fig:exp}\,(c)), we then determine the centroids of the calibration lines and associate them with the theoretical references mentioned above. By fitting a third-order polynomial to these data, we obtain the dispersion curve and thus the calibrated monochromator energy scale.

For the \ion{Fe}{17} measurements, we set the EBIT to use a $\sim4$~mA, 3500\,eV electron beam, capable of directly exciting the lines studied here, and thus generating an undesired background. These parameters yielded a ratio of photoexcitation peak to electron-impact background between 2 and 3 throughout the experiment, indeed lower than the ratio of $\approx$45 achieved in our previous work~\citep{kuhn2022} by cyclically switching the electron-beam energy between ion breeding and probing energy after a long parameter optimization. This time, since switching tests showed a severe loss of \ion{Fe}{17} ions, we decided to use a constant electron-beam energy of 3500\,eV, well above that of dielectronic recombination satellites. The present signal-to-noise ratio and resolving power of 13,000 were sufficient for our reported accuracy.

The P04 monochromator was scanned over ranges covering $3s-2p$ (3G and 3F), $3d-2p$ (3C and 3D), $3p-2s$ (3A and 3B), and $4d-2p$ transitions (4C and 4D) of \ion{Fe}{17}. Fluorescence was collected in the SDDs for 10--15 s at each monochromator step. The count rate for each transition is directly proportional to the respective oscillator strength, and we can see transitions with excitation rates about 4--80 times lower than that of the 3C transition. Scans of each line were therefore repeated as needed to obtain good statistics. This also yielded adequate statistics at each step for the Au $4f_{7/2}$ photoelectron peak position determination needed for the nominal monochromator photon-energy scale correction. To construct the spectrum for a single transition, as depicted in Fig.~\ref{fig:sp}, all photons detected in the SDDs within a 50-eV region of interest centered around the expected energy are summed as a function of the monochromator energy. A representative scan for each of these lines is shown in Fig.~\ref{fig:sp}. The transition energies of the \ion{Fe}{17} lines were determined using a maximum-likelihood fit of Voigt profiles added to the linear background term arising from electron-impact excitation using the cash statistic~\citep{cash1979,kaastra2017}. The Voigt function is a convolution of Lorentzian and Gaussian functions. The Gaussian contributions to the linewidth arise from the limited resolution of the monochromator and the thermal motion of the ions~\citep{hoesch2022}. The Lorentzian width, as shown in~\cite{kuhn2022}, stems from the natural linewidth of the transition and a pseudo-Lorentz instrumental component due to X-ray diffraction at beamline components~\citep{follath2010}. Given the possible energy-dependent contributions to line widths from beamline components, we chose to leave all parameters of the Voigt profile unconstrained during our fitting procedure for determining the line centroids.

\begin{table*}
  \centering
  \caption{Experimental and calculated transition energies from this work in comparison with previous experiments, astrophysical observations, and other predictions, all in electronvolt (eV). Values in parentheses following the measured values give total uncertainties and parentheses below the measured values indicate absolute differences from the present measurements. }
    \begin{tabular}{@{\extracolsep{\fill}} cccccccccc}
\hline\hline
Line  & Term  & Configuration & \multicolumn{3}{c}{This Work} & \multicolumn{2}{c}{Previous Experiments} & \multicolumn{2}{c}{Observations} \\
      &       &       & Experiment & Large CI$^{a}$ & FAC-MBPT$^{b}$ & BW94$^{c}$ & B98$^{d}$ & Hinode$^{e}$ & SMM$^{f}$ \\
\hline
3G    & $^{3}P_{1}$ & $ [1s^2 2s_{1/2}^2 2p_{1/2}^2 2p_{3/2}^3 3s_{1/2}]_1 $ & 727.073(15) & 727.086 & 727.084 & 727.01(4) & 727.14(4) & 727.14 & 727.14 \\
      &       &       &       & (-0.013) & (-0.011) & (0.06) & (-0.06) & (-0.06) & (-0.06) \\
3F    & $^{1}P_{1}$ & $ [1s^2 2s_{1/2}^2 2p_{1/2} 2p_{3/2}^4 3s_{1/2}]_1 $ & 739.067(15) & 739.034 & 739.002 & 739.23(13) & 738.88(9) & 739.07 & 739.10 \\
      &       &       &       & (0.033) & (0.065) & (-0.17) & (0.19) & (-0.01) & (-0.03) \\
3D    & $^{3}D_{1}$ & $ [1s^2 2s_{1/2}^2 2p_{1/2}^2 2p_{3/2}^3 3d_{5/2}]_1 $ & 812.417(13) & 812.418 & 812.363 & 812.21(11) & 812.43(11) & 812.37 & 812.74 \\
      &       &       &       & (-0.001) & (0.054) & (0.21) & (-0.01) & (0.05) & (-0.33) \\
3C    & $^{1}P_{1}$ & $ [1s^2 2s_{1/2}^2 2p_{1/2} 2p_{3/2}^4 3d_{3/2}]_1 $ & 825.870(12) & 825.852 & 825.765 & 826.07(6) & 825.79(6) & 825.85 & 825.90 \\
      &       &       &       & (0.019) & (0.106) & (-0.20) & (0.08) & (0.02) & (-0.03) \\
3B    & $^{3}P_{1}$ & $ [1s^2 2s_{1/2} 2p_{1/2}^2 2p_{3/2}^6 3p_{1/2}]_1 $ & 892.496(10) & 892.490 & 892.446 & 892.68(13) & 892.49(19) & 892.61 & 892.61 \\
      &       &       &       & (0.007) & (0.050) & (-0.18) & (0.01) & (-0.12) & (-0.12) \\
3A    & $^{1}P_{1}$ & $ [1s^2 2s_{1/2} 2p_{1/2}^2 2p_{3/2}^6 3p_{3/2}]_1 $ & 896.774(10) & 896.770 & 896.752 & 896.81(6) & 896.81(13) & 896.81 & 896.88 \\
      &       &       &       & (0.004) & (0.022) & (-0.04) & (-0.04) & (-0.04) & (-0.10) \\
4D    & $^{3}D_{1}$ & $ [1s^2 2s_{1/2}^2 2p_{1/2}^2 2p_{3/2}^3 4d_{5/2}]_1 $ & 1010.983(16) & -     & 1010.921 &   -    & 1010.80(8) & 1010.96 & 1011.04 \\
      &       &       &       &       & (0.062) &       & (0.19) & (0.02) & (-0.06) \\
4C    & $^{1}P_{1}$ & $ [1s^2 2s_{1/2}^2 2p_{1/2} 2p_{3/2}^4 4d_{3/2}]_1 $ & 1022.639(16) & -     & 1022.552 &    -   & 1022.63(8) & 1022.62 & 1022.80 \\
      &       &       &       &       & (0.087) &       & (0.01) & (0.02) & (-0.16) \\
 \hline\hline
\end{tabular}%
{\begin{flushleft}
\footnotesize
\tablenotemark{$^a$}{Large CI calculations, method is from ~\cite{cheung2021scalable}}\\
\tablenotemark{$^b$}{CI + Second-order MBPT of FAC, method is from~\cite{gu2006inner}}\\
\tablenotemark{$^c$}{EBIT measurements by \cite{wargelin1994}}\\
\tablenotemark{$^d$}{EBIT measurements by \cite{brown1998laboratory}}\\
\tablenotemark{$^e$}{Solar observations by \textit{Hinode}: \cite{delzanna2009}}\\
\tablenotemark{$^f$}{Solar observations by \textit{SMM}: \cite{phillips1982}}\\
\end{flushleft}
}
  \label{tab:Tab1}%
\end{table*}%
 
\begin{table*}[htbp]
  \centering
  \caption{Continuation of Table 1. The experimental results are compared with previous predictions, with energy units expressed in electronvolts. Values in parentheses below the predicted values denote the absolute differences from the current measurements.}
\begin{tabular}{cccccccccccc}
    \hline\hline
    Line  & Large CI$^{g}$ & Exp.$^{h}$ & NIST ASD$^{i}$ & AtomDB$^{j}$ & CHIANTI$^{k}$ & CHIANTI$^{l}$ & SPEX$^{m}$ & W16$^{n}$ & S15$^{o}$ & G05$^{p}$ & A04$^{q}$ \\
          &   (old)    &       &       & CI    & MRMP  & AS    &       & MBPT  & MBPT  & MBPT  & MCDF \\
    \hline
3G    & 726.97 & 727.11 & 727.14 & 725.79 & 727.06 & 727.48 & 727.18 & 726.78 & -     & 727.12 & 725.38 \\
      & (0.10) & (-0.03) & (-0.07) & (1.28) & (0.01) & (-0.41) & (-0.11) & (0.29) &       & (-0.05) & (1.70) \\
3F    & 738.91 & 739.04 & 739.05 & 738.01 & 739.00 & 738.21 & 738.88 & 738.72 & -     & 739.06 & 736.05 \\
      & (0.16) & (0.03) & (0.01) & (1.06) & (0.07) & (0.85) & (0.19) & (0.34) &       & (0.01) & (3.02) \\
3D    & 812.32 & 812.41 & 812.37 & 811.70 & 812.41 & 813.65 & 812.48 & 812.04 & 812.57 & 812.44 & 811.08 \\
      & (0.10) & (0.01) & (0.05) & (0.72) & (0.01) & (-1.23) & (-0.06) & (0.37) & (-0.15) & (-0.02) & (1.34) \\
3C    & 825.76 & 825.83 & 825.70 & 825.83 & 825.76 & 827.52 & 826.01 & 825.39 & 825.89 & 825.70 & 825.01 \\
      & (0.11) & (0.04) & (0.17) & (0.04) & (0.11) & (-1.65) & (-0.14) & (0.48) & (-0.02) & (0.17) & (0.86) \\
3B    & -     & -     & 892.50 & 894.25 & 892.41 & 895.55 & 892.61 & 892.21 & -     & 892.40 & 894.25 \\
      &       &       & (-0.00) & (-1.75) & (0.08) & (-3.05) & (-0.12) & (0.29) &       & (0.10) & (-1.75) \\
3A    & -     & -     & 896.90 & 898.54 & 896.67 & 899.85 & 897.14 & 896.46 & -     & 896.62 & 898.55 \\
      &       &       & (-0.13) & (-1.77) & (0.10) & (-3.08) & (-0.36) & (0.31) &       & (0.15) & (-1.77) \\
4D    & -     & -     & 1011.00 & 1009.79 & -     & 1012.03 & 1011.29 & 1010.53 & -     & -     & 1009.22 \\
      &       &       & (-0.02) & (1.19) &       & (-1.05) & (-0.31) & (0.46) &       &       & (1.76) \\
4C    & -     & -     & 1022.70 & 1021.76 & -     & 1023.64 & 1022.97 & 1022.17 & -     & -     & 1020.90 \\
      &       &       & (-0.06) & (0.88) &       & (-1.00) & (-0.33) & (0.47) &       &       & (1.74) \\
   \hline\hline
    \end{tabular}%
{\begin{flushleft}
\footnotesize
\tablenotemark{$^g$}{Large CI: \cite{kuhn2022}}\\
\tablenotemark{$^h$}{Preliminary critical analysis of Fe XVII spectral data, Private Communication \cite{AK} }\\
\tablenotemark{$^i$}{NIST Atomic Spectroscopy Database: \cite{NIST_ASD}}\\
\tablenotemark{$^j$}{AtomDB Database: \cite{lpb2006} (APED: \texttt{fe\_17\_LV\_v3\_0\_4\_a.fits)}}\\
\tablenotemark{$^k$}{Chianti Database with MRMP calculations: \cite{delzanna2009}}\\
\tablenotemark{$^l$}{Chianti Database with Autostructure (AS) calculations: \cite{liang2010}}\\
\tablenotemark{$^m$}{SPEX database: \cite{gu2020capellaEBIT}}\\
\tablenotemark{$^n$}{MBPT by \cite{wang2016}}\\
\tablenotemark{$^o$}{MBPT by \cite{santana2015electron}}\\
\tablenotemark{$^p$}{MBPT by \cite{gu2005}}\\
\tablenotemark{$^q$}{MCDF by \cite{aggarwal2003}}\\
\end{flushleft}}
  \label{tab:Tab2}%
\end{table*}%

Table~\ref{tab:Tab1} presents the results for eight \ion{Fe}{17} lines and their associated uncertainties from errors in the centroid determination of calibration lines, the dispersion fit 1$\sigma$ confidence band, and the centroid determination of each \ion{Fe}{17} line, which is typically in the range of 1--3 meV. The total systematic uncertainties of the calibration are estimated to reach levels of 10--15 meV. As mentioned before, the angular encoder interpolation error induces oscillations of the nominal monochromator photon energy scale up to $\pm70$~meV in the 650--1150 eV energy range. While accurate reference energies and the corrections from ASPHERE to the photon-energy axis help mitigate these oscillations, $\sim$10--20\% (7-14 meV) residual variations still remain in the corrected monochromator photon-energy scale. A potential source of these could be the limited resolution of the Keithley 6517 voltage source biasing the gold target. Despite using a seven-digit calibrated voltmeter (Agilent 3458a), the voltage source operates in 5\,mV steps within the 100 V range, limiting our electron kinetic energy measurements. Further systematics arise from the frequent switching of the voltage range of this bias supply needed to cover the monochromator range of 600--1150 eV, requiring separate calibration for each voltage range. Unfortunately, we could only calibrate the bias supply in a narrow 20 V range. Moreover, unmeasured fluctuations in the voltages applied to the inner and outer hemispheres of the electron spectrometer may have introduced additional systematic uncertainties. The dimensional stability of its electron-optics components is affected by thermal drifts caused by varying ambient conditions at ppm levels to which we are already sensitive. Note that we also fit our data with skewed Voigt profiles, allowing for a nonzero skewness, thus accounting for any line asymmetries that may exist due to monochromator imperfections~\citep{perry-Sassmannshausen2021,hoesch2022,togawa2023}. These tests resulted in changes to the line centroids of less than $\sim1$~meV, which is negligible considering the total uncertainty of our measurements. We considered whether lines from contaminant ions of oxygen, fluorine, or neon originating from residual calibration gases could lead to systematic errors in any of our transition-energy determinations. Because these lines have known transition energies, and because of the extremely high resolving power (13,000) attained in our experiment, we ruled out any significant effect from such contamination. After conservatively considering all these sources, our present uncertainties are a factor of 4--20 smaller than those of previously reported experiments~\citep{wargelin1994,brown1998laboratory}.

\section{Discussion of the results}
\label{result}

\begin{table*}[htbp]
  \centering
  \caption{\label{tabT} Contributions to the theoretical energies (in cm$^{-1}$ above the ground state) of \ion{Fe}{17} from an enlarged basis set ($>17g$), additional reference configurations (Extras) and QED in comparison with our measurements and their errors in cm$^{-1}$. Note that we used the CODATA2018~\citep{codata2018} recommended value of $hc$ to convert experimental values from eV to cm$^{-1}$. The difference between
  	the three pairs of lines is shown in bold.}
    \begin{tabular}{ccccccccccc}
    \hline\hline
    Label & This Exp. & Error & $\Delta$Prev. Th.$^a$ & $17g$ & $\ge 17g$ & Extras & QED   & Final & $\Delta$This Th.$^b$ & $\Delta$This Th.$^b$ (\%)\\
    \hline
    3G    & 5864241 & 122   & 841   & 5862842	 & 541   & 146   & 814   & 5864343 & --102  & 0.0017 \\
    3F    & 5960976 & 123   & 1274  & 5958941	 & 558   & 146   & 1067  & 5960711 & 265   & 0.0045 \\
    \textbf{3F--3G} & \textbf{96736} & \textbf{174} & \textbf{434} & \textbf{96099}	 & \textbf{17} & \textbf{0} & \textbf{253} & \textbf{96368} & \textbf{368} & \textbf{0.3800}\\
    \hline
    3D    & 6552587 & 103   & 787   & 6552044	 & 294   & 104   & 151   & 6552594 & --7    & 0.0001\\
    3C    & 6661093 & 96    & 897   &   6660390	 & 248   & 5     & 299   & 6660942 & 151   & 0.0023 \\
    \textbf{3C--3D} & \textbf{108506} & \textbf{141} & \textbf{110} &   \textbf{108346}	 & \textbf{--46} & \textbf{--99} & \textbf{148} & \textbf{108348} & \textbf{158} & \textbf{0.1454}\\
    \hline
    3B    & 7198469 & 82    &       & 7200865	 & 573   & --28   & --2993 & 7198416 & 53    & 0.0007\\
    3A    & 7232969 & 82    &       & 7235357	 & 547   & --8    & --2958 & 7232938 & 31    & 0.0004 \\
    \textbf{3A--3B} & \textbf{34500} & \textbf{116} &       & 	\textbf{34492}	 & \textbf{--25} & \textbf{20} & \textbf{35} & \textbf{34522} & \textbf{--22} & \textbf{0.0626}\\
    \hline\hline
    \end{tabular}%
\begin{flushleft}
$^a$This column shows the difference between previous theoretical large CI computations from \cite{kuhn2022} and those of the present experiment.\\
$^b$This column shows the difference between current theoretical large CI computations and those of the present experiment.
\end{flushleft}
\end{table*}%

We compare the present results in Fig.~\ref{fig:sp} and Tabs.~\ref{tab:Tab1} and~\ref{tab:Tab2} with earlier experimental data, observations, and predictions, including our own. Our calculations employ the latest version of our highly scalable parallel CI code \citep{cheung2021scalable} (see Appendix). Optimization of the basis-set construction allowed faster convergence with the principal quantum number $n$ than in our prior work (see supplementary material of \cite{kuhn2022}), while including higher partial waves ($h$, $i$, and $k$), and a larger number of reference configurations of even and odd parity. Table~\ref{tabT} shows the QED and other contributions in cm$^{-1}$ for the measured transitions. Column $17g$ shows the results obtained with the $17spdfg$ basis set (see the Appendix); column $>17g$, additional contributions from highly excited orbitals up to $24spdfgh21i17k$; ``Extras'', additional contributions due to the much larger number of configurations included in CI, selected to give the large contributions. The final results are the sum of these three columns and a QED contribution~\citep{tupitsyn2016}. Columns [$\Delta$This Th.$^b$] show differences between the current experiment and theory, and [$\Delta$Prev. Th.$^a$] show differences from previous calculations presented in \citet{kuhn2022}, demonstrating a significant improvement. We estimate the uncertainty in the electronic correlation computations to be approximately 29 meV ($\sim$230~cm$^{-1}$), primarily arising from the $>17g$ contribution (see Appendix). The difference between theory and experiment is within the combined uncertainties for all six levels. This allows us to estimate the uncertainty of the QED contribution at 30--32 meV (240--260 cm$^{-1}$), which is the combined theory and experimental uncertainties added in quadrature. We also computed the energies of the 3A and 3B levels for the first time. The $2s-3p$ lines (3A and 3B), which involve a $2s$ electron, have the largest QED contributions ($\sim370$ meV, or $\sim3000$ cm$^{-1}$), while for $3s-2p$ transitions 3G and 3F as well as $3d-2p$ ones (3C and 3D) they are much smaller. From the 3A and 3B results, we thus estimate a relative QED accuracy of 8\%. 

As shown in Tab.~\ref{tab:Tab1}, line 3F, close to the He-like F~K$\alpha$ calibration line, shows a larger absolute deviation of about 33 meV from the large CI prediction, while the remaining measured lines remain below $\sim$10--20 meV. Unfortunately, line 3F was measured only once, unlike the others, which were scanned at least 4--5 times. We explored several plausible explanations for the 3F discrepancy. One possible source of the discrepancy could be the simultaneous excitation of high-$n$ Rydberg lines of \ion{O}{7} within the scan range of line 3F, which could lead to a shift of the 3F centroid. Furthermore, we considered lines from the lower charge states, \ion{Fe}{10}, \ion{Fe}{8}, and \ion{Fe}{7}, which fall within the 3F scan range. Despite the relatively low abundance of these charge states in our experiment, they can potentially influence the 3F line due to their strong oscillator strengths. Although the theoretical line positions and oscillator strengths of these low charge states are calculated by \citet{gu2006inner}, they have never been compared experimentally, making it difficult to estimate their influence on the 3F position. We also investigated the possibility of magnetically induced mixing of the $J=0$ and $J=1$ $(2p_{1/2}^{-1}\,3s_{1/2})$ excited states~\citep{beiersdorfer2003MIT}, which might shift the energy of the $J=1$ state sufficiently to introduce a systematic error in our measurement of 3F. However, measurements by~\cite{beiersdorfer2016MIT} show a separation of $\sim$1.2 eV between these states, making strong magnetic-field-induced mixing unlikely. We performed FAC calculations for atoms in strong magnetic fields to verify this, finding shifts on the order of 10~\textmu eV for the field strength in PolarX-EBIT, demonstrating that this effect is not important in our experiment. The decrease in reflectivity of the platinum-coated diffraction grating over the 3F scan range could slightly affect the centroid position determination at 739 eV. Based on simulations we estimate this effect to be smaller than 0.1 meV.

We also consider the differences between the three line pairs, as they are more sensitive to QED effects than absolute energies. Table~\ref{tabT} shows that the largest uncertainty, caused by the uncertainty in the basis-set convergence, is common to each of the pairs. This significantly reduces the uncertainty of electronic correlations to better than 6 meV (50~cm$^{-1}$) for the energy difference. Both (3A--3B) and (3C--3D) are in excellent agreement with our present as well as previous predictions~\citep{kuhn2022}. For (3G--3F), the deviation is 46 meV (about $2\sigma$) and can be attributed to the factors discussed above for the line 3F. It is interesting to note that our measured 3F energy is in much better agreement with solar observations~\citep{phillips1982,delzanna2009} than with our calculations. Nevertheless, our present calculations of the ground state transitions show an order of magnitude smaller deviation from our experimental results compared to our prior predictions~\citep{kuhn2022}. This represents a benchmark with our experimental data at the level of 10--20 ppm, an unprecedented agreement for a neon-like system to the best of our knowledge.

Besides CI, we performed calculations using a combination of conventional CI and second-order many-body perturbation theory (MBPT) with the Flexible Atomic Code (FAC)~\citep{gu2008flexible}. Details of this method are described in~\cite{gu2005,gu2006inner}, and recently in~\cite{steinbruegge2022}. In these calculations, we included frequency-dependent generalized Breit interactions in both the CI expansion and the MBPT corrections, as well as self-energy and vacuum polarization calculated using the QED operator model of~\cite{shabaev2018}. These predictions demonstrate a reasonable agreement with our experimental data, with the largest discrepancy of about 100~meV observed for line 3C. We compared our results with other CI+MBPT data available in the literature~\citep{wang2016,santana2015electron,gu2005}, showing maximum deviations of up to 0.5~eV. The origin of the discrepancy between our CI+MBPT calculations and the previously published ones is unclear. We have also observed departures from the predictions of multiconfiguration Dirac-Fock (MCDF) and autostructure (AS) calculations~\citep{aggarwal2003,loch2005effects,liang2010}, with deviations reaching up to 1--3 eV. However, we note that the atomic structures used in these calculations were necessarily small to facilitate their use in R-matrix collision calculations, which are computationally more demanding compared to atomic structure calculations. Other accurate predictions for \ion{Fe}{17} from multireference Møller-Plesset (MRMP) are reported in \cite{delzanna2009} and included in the CHIANTI code. They show very good agreement with our experimental data. 

We compare our results with laboratory data from \citet{wargelin1994} and \citet{brown1998laboratory}. Both works measured electron-impact spectra of Fe XVII under similar experimental conditions in the Lawrence Livermore National Laboratory (LLNL) EBIT facility using a crystal spectrometer employing a cesium acid phthalate (CsAP) crystal for the wavelength range of the lines presently discussed. Both measurements have carefully concatenated several spectra from different wavelength ranges and calibrated them against reference lines of hydrogenic and heliumlike oxygen, fluorine, and neon, similar to our work. Nevertheless, these two measurements are themselves marginally inconsistent with each other within their quoted uncertainties. Furthermore, we find that our measurements are also marginally inconsistent with both these previous measurements within uncertainties. The source of these marginal inconsistencies is unknown.

We also compare our results with data from widely used databases and plasma codes. The NIST Atomic Spectroscopy Database (ASD)~\citep{NIST_ASD} values showed significant deviations for lines 3C and 3A. However, when critically evaluated $n=3-2$ data by the authors of the NIST ASD~\citep{AK} were considered, we found a much better agreement with our experimental results (see Tab.~\ref{tab:Tab2}). Comparison with AtomDB \citep{foster2012}, CHIANTI \citep{chianti2021}, and SPEX \citep{kaastra1996spex} databases and plasma codes revealed discrepancies as large as 1--2 eV. SPEX numbers showed better agreement with our results than those found in AtomDB, since SPEX has updated Fe-L atomic data \citep{gu2019, gu2020capellaEBIT, gu2022}, which were mainly calculated using FAC. Although the Astrophysical Plasma Emission Database (APED) version in AtomDB shows different values in its online webguide version (2.0.1)~\footnote{\url{http://www.atomdb.org/Webguide/webguide.php}} and its pyatomdb version (3.0.4), the theoretical source is in both cases~\cite{loch2005effects}, which uses the CI method, and disagrees by up to 1--2 eV from our results, as shown in Tab.~\ref{tab:Tab2}. CHIANTI provides two sources for the energies: Autostructure (AS) theory \citep{liang2010}, and the more accurate set of data from MRMP theory \citep{delzanna2009}.
Note that the theoretical energy level data in AtomDB and Chianti are not used when generating model spectra when more accurate experimental or observational values exist. AtomDB replaces the most important transition energies of Fe XVII with the \cite{brown1998laboratory} values, whereas CHIANTI uses the observed transition energies from solar Observation~\citep{delzanna2009}.

Overall, most experimental and observational data agree with our experiment within 0.1 eV on average, well within the error bars of earlier works. However, there are substantial discrepancies with predictions from certain theoretical models, exceeding the margins of error associated with the experimental results. This highlights the urgent need to update the aforementioned databases to avoid pitfalls in astrophysical spectrum modeling and interpretation of observational data.

\section{Summary and Conclusions}\label{concl}

We presented high-precision transition-energy measurements of eight strong, astrophysically preeminent \ion{Fe}{17} transitions required for plasma diagnostics. Our approach combined resonant photoexcitation of \ion{Fe}{17} and narrow H-like and He-like transitions with high-resolution photoelectron spectroscopy ~\citep{ROSSNAGEL2001}. This eliminates a very common source of systematic errors found even in advanced monochromators, namely quasiperiodic encoder interpolation errors~\citep{follath2010,Krempasky2011}. As a result, our \ion{Fe}{17} measurements represent a significant improvement in accuracy compared to previous experimental references, achieving an average enhancement of almost an order of magnitude. The uncertainties now stand at 10--15 meV, which translates to Doppler shifts of approximately $\pm5$~km\,s$^{-1}$. A further improvement in accuracy by another order of magnitude will require incorporating high-resolution/-stability voltage sources and more accurate voltmeters at ASPHERE to eliminate systematic errors associated with knowledge of the bias voltages.

We have also improved our high-precision calculations by an order of magnitude in comparison with previous best calculations~\citep{kuhn2022,cheung2021scalable}. This improvement allowed us, for the first time, to test the accuracy of QED corrections to the transition energies of a complicated 10 electron system. We expect that the achieved QED accuracy is applicable to a broad range of ions of intermediate degrees of ionization that can be treated with our large-scale CI or CI+all-order approaches. This has significant implications for predicting energy levels in systems where no experimental data are available for a wide range of applications in astrophysics, plasma physics, and atomic clock development \citep{Piet}. The established QED accuracy is deemed sufficient for high-precision prediction of HCI clock transitions \citep{HCI}. Improved accuracy of the experimental values would allow us to further decouple the uncertainty due to basis-set convergence from the uncertainty in the QED and improve theory predictions.

Our improved transition-energy measurements for \ion{Fe}{17} are sufficiently accurate that the uncertainties are no longer a significant part of the error budget for present or future planned astrophysical instruments, such as \chandra HETGS, \xmm RGS, \textit{XRISM}~\citep{xrism2018}, \textit{Athena}~\citep{pajot2018athena}, \textit{LEM}~\citep{lem2022}, \textit{HUBS}~\citep{hubs2020}, \textit{Arcus}~\citep{arcus_instrum}, \textit{HiReX}~\citep{hirex2021}, and \textit{Lynx}~\citep{Schwartz2019}. Future campaigns of similar measurements of prominent transitions in key ions (especially Fe-L shell ions) would be of great utility and could easily be directly included in commonly used astrophysical plasma spectral databases.

The closeness of our large CI calculations to our measured values, with the worst deviation at line 3F of 33 meV amounting to a Doppler shift of only 13 km\,s$^{-1}$, shows that such well-converged calculations are sufficiently accurate to be readily used in spectral databases. While there is no reason for this in the case of the lines measured in the present work, when accurate measurements are not available, similarly well-converged results could be used for other transitions of \ion{Fe}{17} and many other ions. By including a very large number of configurations, our agreement becomes significantly better than that of other well-performing methods, such as results from less-converged large CI, MBPT, and MRMP calculations. For \ion{Fe}{17}, our calculations are more accurate than even the best measurements for Fe-L shell transitions in Li-like through F-like ions \citep{Brown2002}. This suggests a near-future research program composed of comprehensive large CI calculations of transition energies for all ions of astrophysical interest up to Ne-like, supplemented by targeted experiments aimed at measuring the most important transition energies.

\begin{acknowledgments}
This research was funded by the Max Planck Society (MPG) and the German Federal Ministry of Education and Research (BMBF) under project 05K13SJ2. C.S. acknowledge supports from NASA under grant number 80GSFC21M0002 and MPG. F.S.P. and M.A.L. acknowledge support from the NASA Astrophysics Program. The theoretical work has been supported by the US NSF Grants No. PHY-2012068, PHY-2309254 and US Office of Naval Research Grant No. N00014-20-1-2513. Calculations were performed in part using the computing resources at the University of Delaware, in particular the Caviness and DARWIN high-performance computing clusters. M.S.S. thanks MPIK for hospitality. We thank DESY (Hamburg, Germany), a member of the Helmholtz Association HGF, for the provision of experimental facilities. Parts of this research were carried out at PETRA~III. We thank the P04 team at PETRA~III for their skillful and reliable work. We also thank the anonymous referees whose comments and suggestions helped improve and clarify this manuscript.
\end{acknowledgments}

\appendix

\section{Large-scale CI calculations}

In this work, we conducted extensive high-precision calculations of \ion{Fe}{17}. 
We start from the solution of the Dirac-Hartree-Fock equations in the central field approximation to construct the one-particle orbitals. Calculations are carried out using a CI method, correlating all 10 electrons. Breit interaction is included in all calculations. QED corrections are taken from the previous work \citep{kuhn2022} except for the levels with a $2s$ hole, which were not computed in 2022. The same method is used in all QED calculations \citep{tupitsyn2016}. The CI wave function is obtained as a linear combination of all distinct states of a given angular momentum $J$ and parity:
\begin{equation}\label{eq:conf}
\Psi_J = \sum_i c_i\Phi_i.
\end{equation}

The low-lying energies and wave functions are determined by solving the time-independent multielectron Schr\"odinger equation
\begin{equation}
H \Phi_n = E_n \Phi_n.
\end{equation}

Expanding the previous work~\citep{kuhn2022}, we perform several calculations optimizing the basis-set convergence, including higher partial waves up to $k$ orbitals, and significantly expanding the set of reference configurations until convergence is reached in these parameters as well. 

We have shown a comparison of our theoretical results for six transitions measured in this work with the experiment in the main text. We note that such a larger-scale computation of the 4C and 4D levels is beyond the capabilities of available computational resources (32 TB of memory and about 2000 CPUs on our largest available partition). Computing higher-lying levels requires computing of all the lower-lying levels with the same angular momentum and parity, drastically increasing memory requirement. 

To test the consistency of our approach, we compare the data for the even and larger number of odd levels in Tab.~\ref{fe_table1} with the preliminary critical analysis of Fe XVII spectral data by \cite{AK}. These data  generally agree well with our experiment (except for levels with a $2s$ hole), so they serve as a good general reference for other levels. The six levels measured in this work are shown in bold.  The final values are given in cm$^{-1}$ in Tab.~\ref{fe_table1} in the column ``Final''. The difference between the final values and the experimental values in cm$^{-1}$ and percentage are given in the last two columns.

We will discuss a complete assessment of the main contributions to the energies, including the basis-set construction, the inclusion of extra configurations, and QED.  We find excellent agreement with experiments for all energies, at the level of 0.0004\% for some levels. With the high level of accuracy attained, we are able to test QED contributions in the calculations of multielectron systems for the first time.

\begin{table*}[t]
\scriptsize
\caption{\label{fe_table1} Contributions to \ion{Fe}{17} energies calculated with increased basis sets and number of configurations. The results are compared with the preliminary critical analysis of Fe XVII spectral data by \cite{AK}. All energies are given in cm$^{-1}$. The basis set is designated by the highest principal quantum number and the highest partial wave included. For example, $17g$ means that all orbitals up to $n=17$ are included for $spdfg$ partial waves. The last two columns show the differences of the present computations with \cite{AK} in cm$^{-1}$ and \%, respectively. }
\begin{ruledtabular}
\begin{tabular}{lccccccccccccccccc}
\multicolumn{2}{c}{Configuration}&
\multicolumn{1}{c}{Expt$^a$}&
\multicolumn{1}{c}{$\Delta$}&
\multicolumn{1}{c}{$17g$}&
\multicolumn{1}{c}{$+20g$}&
\multicolumn{1}{c}{$+24g$}&
\multicolumn{1}{c}{$+17h$}&
\multicolumn{1}{c}{$+20h$}&
\multicolumn{1}{c}{$+24h$}&
\multicolumn{1}{c}{$+17i$}&
\multicolumn{1}{c}{$+21i$}&
\multicolumn{1}{c}{$+17k$}&
\multicolumn{1}{c}{QED}&
\multicolumn{1}{c}{Extras}&
\multicolumn{1}{c}{Final}&
\multicolumn{1}{c}{$\Delta$}&
\multicolumn{1}{c}{$\Delta$(\%)} \\
\multicolumn{3}{c}{}&
\multicolumn{1}{c}{Ref.$^b$}&
\multicolumn{12}{c}{}&
\multicolumn{2}{c}{Present}\\
\hline
$2s^2 2p^6   $  & $^1S_0$    &   0       &        0 &0    &  0 & 0  &   0 &  0 &  0 & 0   & 0   & 0   & 0     &  0  & 0        & 0   &   0    \\
$2s^2 2p^5 3p$  & $^3S_1$    &   6093295 &  1124&6092365 &  44 & 20 & 278 & 58 & 45 & 64  & 86  & 8   & 70    & 107 & 6093143  & 152 & 0.002\%  \\
$2s^2 2p^5 3p$  & $^3D_2$    &   6121484 & 988 & 6120688 &  38 & 18 & 252 & 51 & 40 & 56  & 77  & 4   & 56    &     & 6121280  & 204 & 0.003\%  \\
$2s^2 2p^5 3p$  & $^3D_3$    &   6134539 &1015&  6133678 &  41 & 19 & 258 & 54 & 42 & 58  & 81  & 5   & 107   &     & 6134345  & 194 & 0.003\%  \\
$2s^2 2p^5 3p$  & $^1P_1$    &   6143639 & 1013&  6142785 & 39 & 18 & 253 & 52 & 40 & 56  & 78  & 4   & 93    &     & 6143417  & 222 & 0.004\%  \\ [0.5pc]
$2s^2 2p^5 3s $ & $  2    $  &   5849216 &1134&  5847527 &  38 & 16 & 269 & 52 & 35 & 63  & 81  & 8   & 813   & 149 & 5849052  & 164 & 0.003\%  \\
{$\mathbf{2s^2 2p^5 3s}$} & {$\mathbf{^{3}P_{1}}$}  &   \textbf{5864502} & \textbf{1102}& \textbf{5862842} &  \textbf{37} & \textbf{15} & \textbf{258} & \textbf{50} & \textbf{34} & \textbf{60} & \textbf{81}  & \textbf{7} & \textbf{814} & \textbf{146} & \textbf{5864343}  & \textbf{158} & \textbf{0.003\%}  \\
{$\mathbf{2s^2 2p^5 3s}$} & {$\mathbf{^{1}P_{1}}$}  &   \textbf{5960742} &\textbf{1040}&   \textbf{5958941} & \textbf{41} & \textbf{18} & \textbf{259} & \textbf{55} & \textbf{37} & \textbf{60} & \textbf{81}  & \textbf{7} & \textbf{1067} & \textbf{146} & \textbf{5960711}  & \textbf{31} & \textbf{0.001\%}  \\
$2s^2 2p^5 3d $ & $^3P_1^o$  &   6471640 &1148&  6470765 &  51 & 24 & 138 & 65 & 47 & 22  & 81  & -2  & 95   &  139  & 6471426  & 214 & 0.003\%  \\
$2s^2 2p^5 3d $ & $^3P_2^o$  &   6486183 &1007&   6485436 & 51 & 24 & 121 & 65 & 47 & 17  & 81  & -4  & 109  &  139  & 6486086  &  97 & 0.001\%  \\
$2s^2 2p^5 3d $ & $^3F_4^o$  &   6486720 &920 &  6486064 &  51 & 24 &  90 & 65 & 47 & 7   & 81  & -7  & 105  &  142  & 6486669  &  51 & 0.001\%  \\
$2s^2 2p^5 3d $ & $^3F_3^o$  &   6492651 &856&  6492060 &  50 & 23 &  66 & 64 & 46 & -2  & 81  & -10 & 102  &  138  & 6492621  &  30 & 0.000\%  \\
$2s^2 2p^5 3d $ & $^1D_2^o$  &   6506537 &855&   6505941 & 50 & 23 &  62 & 64 & 46 & -3  & 81  & -10 & 107  &  138  & 6506500  &  37 & 0.001\%  \\
$2s^2 2p^5 3d $ & $^3D_3^o$  &   6515203 &  807&6514654 &  50 & 23 &  49 & 63 & 46 & -8  & 81  & -12 & 107  &  136  & 6515189  &  14 & 0.000\%  \\
{$\mathbf{2s^2 2p^5 3d}$} & {$\mathbf{^3D_1^o}$}  &   \textbf{6552503} & \textbf{703}& \textbf{6552044} &  \textbf{51} & \textbf{24} & \textbf{ 49} & \textbf{65} & \textbf{47} & \textbf{-9} & \textbf{81}  & \textbf{-12} & \textbf{151} & \textbf{104} & \textbf{6552594}  & \textbf{91} & \textbf{0.001\%}  \\
$2s^2 2p^5 3d $ & $^3F_2^o$  &   6594309 &  802&6593569 &  55 & 26 &  71 & 69 & 50 & 0   & 81  & -9  & 355  &  138  & 6594404  & 95  & 0.001\%  \\
$2s^2 2p^5 3d $ & $^3D_2^o$  &   6600998 &938&   6600124 & 54 & 26 &  80 & 69 & 49 & 2   & 81  & -8  & 349  &  137  & 6600962  & 36 & 0.001\%  \\
$2s^2 2p^5 3d $ & $^1F_3^o$  &   6605185 & 857& 6604381 &  54 & 26 &  56 & 69 & 49 & -6  & 81  & -11 & 363  &  136  & 6605198  & 13 & 0.000\%  \\
{$\mathbf{2s^2 2p^5 3d}$} & {$\mathbf{^1P_1^o}$}  &   \textbf{6660770} & \textbf{574}&   \textbf{6660390} &\textbf{54} & \textbf{26} & \textbf{ 11} & \textbf{68} & \textbf{49} & \textbf{-23} & \textbf{81}  & \textbf{-17} & \textbf{299} & \textbf{5} & \textbf{6660942}  & \textbf{172} & \textbf{0.003\%}  \\
{$\mathbf{2s 2p^6 3p}$} & {$\mathbf{^3P_1^o}$}  &   \textbf{7199200} & & \textbf{7200865} & \textbf{47} & \textbf{30} & \textbf{248} & \textbf{56} & \textbf{38} & \textbf{65 } & \textbf{81}  & \textbf{8  } & \textbf{-2993} & \textbf{-28} & \textbf{7198416}  & \textbf{784} & \textbf{0.011\%}  \\
$2s 2p^6 3p   $ & $^3P_2^o$  &          & &  7219595 & 48 & 31 & 251 & 57 & 39 & 66  & 81  & 8   & -2944 & -36 & 7217197  &  &             \\
{$\mathbf{2s 2p^6 3p}$} & {$\mathbf{^1P_1^o}$}  &   \textbf{7233292} &&  \textbf{7235357} & \textbf{46} & \textbf{30} & \textbf{235} & \textbf{54} & \textbf{35} & \textbf{61 } & \textbf{81}  & \textbf{6  } & \textbf{-2958} & \textbf{-8} & \textbf{7232938}  & \textbf{354} & \textbf{0.005\%}  \\
\end{tabular}
\begin{flushleft}
$^a$\cite{AK}\\
$^b$\cite{kuhn2022}. This column shows the difference between previous theoretical large CI computations from \cite{kuhn2022} and the preliminary critical analysis of Fe XVII spectral data by \cite{AK}.
\end{flushleft}
\end{ruledtabular}
\end{table*}

\textit{Computation}----We consider \ion{Fe}{17} as a system with 10 valence electrons and start with all possible single and double excitations to any orbital up to $17spdfg$ from the $1s^2 2s^2 2p^6$ and $1s^2 2s^2 2p^5 3p$ even-parity reference configurations, and the $1s^2 2s^2 2p^5 3s$, $1s^2 2s^2 2p^5 3d$ and $1s^2 2s 2p^6 3p$ odd-parity reference configurations. For example, a single excitation from the reference configurations $2s^2 2p^6$ can include promoting an electron from the $2s$ or $2p$ orbitals to any orbital up to $17s$, $17p$, ... $17g$, with $2s^2 2p^5 10p$ or $2s 2p^6 17s$ as example outcomes. We designate the basis set by the highest principal quantum number and the highest partial wave included. For example, $17g$ means that all orbitals up to $n=17$ are included for $spdfg$ partial waves. Note that $1s^2$ is removed from all the designations to save space.

The base calculation for the energy levels is done with a $17g$ basis set and is listed in cm$^{-1}$ in Tab.~\ref{fe_table1} in column ``$17g$''. The contributions to the energy levels from expanding the basis set to $20g$ and $24g$ are in the columns ``$+20g$'' and ``$+24g$'', respectively. The largest difference between the $23g$ and $24g$ calculations was 3 cm$^{-1}$, so the basis set at the level of $spdfg$ partial waves is considered sufficiently saturated. We note that although the $24spdfg$ basis was also used in \cite{kuhn2022}, we constructed more compact  basis in the present work, to significantly improve convergences with the principal quantum number $n$. The basis is constructed in the 5 a.u. cavity, while the basis in \cite{kuhn2022} was constructed in a 20 a.u. cavity, with additional differences in the constructions of the higher partial-wave orbitals. A detailed comparison of the two computations confirms much better convergence properties of the present basis. We note very large computational resources needed for a basis-set expansion, especially for the inclusion of higher partial waves. 

Contributions to higher partial waves are considered in the next six columns of Tab.~\ref{fe_table1}. We calculated the contributions of extending the base $17g$ basis set to include up to $17h$ orbitals and listed them in column ``$+17h$''. Next, we successively increase the principal quantum number and increase the basis set up to $24h$. The contributions from $(18-20)h$ orbitals and $(21-24)h$ orbitals are given in columns ``$+20h$'' and ``$+24h$'', respectively. The largest difference between $23h$ and $24h$ calculations was 9 cm$^{-1}$, so the energies of including the higher $h$ orbitals have also converged sufficiently. We note that a large fraction of the $nh$ contribution comes from very high-$n$ orbitals, so the inclusion of the first few $h$ orbitals does not give correct results for this partial wave. This effect is exacerbated for the $i$ and $k$ orbitals, where more of the contribution is expected to come from $n>20$ even with the present compact basis.

The same procedure was used to obtain contributions from the $i$ orbitals up to $21i$ and $k$ orbitals up to $17k$ and are listed under columns ``$+21i$'' and ``$+17k$'', respectively. Contributions from including $i$ orbitals up to the same principal quantum number $n=17$ as the base run are listed in column ``$+17i$. Due to the high computational demand for higher partial-wave calculations, we did not perform calculations for odd-parity states at the level of $21i$. Instead, we set the contributions of $21i$ to the odd-parity energies to be the average of the even-parity state contributions, which was 81 cm$^{-1}$. Contributions from $k$ orbitals were already at a level of convergence around 15 cm$^{-1}$ at $21i$.

We note that we have performed detailed convergence studies computing a separate contribution for each $nl$ for the last few principal quantum numbers to evaluate convergence. Based on these data, we conservatively estimate the missing higher $g$ orbital contribution at 5~cm$^{-1}$, higher $h$ orbital contribution at 20~cm$^{-1}$, and higher $i$ orbital contribution at 50~cm$^{-1}$. It appears that $17k$ is not sufficiently converged. Table~\ref{fe_table1} shows that the contribution of all $ni$ orbitals is about 1/2 of the $nh$ contribution. Conservatively assuming a similar convergence pattern for higher partial waves gives 70~cm$^{-1}$ for the $k$ partial wave and a similar total contribution for all the other partial waves. The total uncertainty due to the convergence of the basis set is then of the order of 230~cm$^{-1}$. However, we note that the incomplete convergence of the basis is expected to cause a systematic shift of data for all levels; i.e., all energy values will be larger, with some smaller variances between the levels. It is possible that the partial-wave convergence is faster and the overall shift is smaller; therefore, we only use the above estimate to make an accuracy evaluation but do not shift the theory values. We note that overall +100~cm$^{-1}$  shift of all of our values would improve the agreement of our data with the present experiment; however, this is the level of the experimental precision at 1~$\sigma$ so improved experimental precision is needed to definitively test the basis set-convergence.

Additionally, an extensive evaluation of the configuration weights was done to include important configurations in the list of basic reference configurations used to construct a final set of configurations. The weights of configurations signify the configuration's contribution to the corresponding wave functions, and are calculated for each configuration $\Phi_i$ as $|c_i|^2$ from Equation~\ref{eq:conf}. These calculations are done by allowing single and double excitations to a much smaller $12g$ basis set since the size of the computational problem will become prohibitive when additional reference configurations are included. The total contributions to including these extra configurations are given in the column ``Extras'' in Tab.~\ref{fe_table1}. Beyond the initial 2 even- and 3 odd-parity configurations, we systematically included an additional 12 even- and 9 odd-parity reference configurations. Note that energies were calculated only for two even-parity levels to save computational resources and allow for additional reference configurations. The inclusion of these extra configurations contributes about 100 cm$^{-1}$ shift to the energies and accounts for an additional 2 million relativistic configurations. Note also that these contributions would also be higher if the calculations were done with a larger basis set. We estimate an uncertainty from the convergence of the CI configuration set at the level of 50~cm$^{-1}$, which is essentially negligible in comparison with the basis-set convergence uncertainty. We note that missing contributions can be  both positive and negative in this case.

Analysis of contributions to the 3F--3G, 3C--3D, and 3A--3B line differences given in the main text Table~\ref{tabT} shows that the basis-set expansion contribution effectively cancels for similar configurations; it is less than 50~cm$^{-1}$ for all three cases. We also find that QED contributions play a major role in the 3F--3G energy difference. For 3C--3D, the contributions from the basis-set expansion and the addition of extra configurations essentially cancel out the QED. In the 3A--3B difference, the basis-set and extra configuration contributions cancel, leaving a shift from the QED. Therefore, comparing the differences in the energy values for similar configurations provides important additional information. It would be very useful to improve the uncertainty of the experiment as well as carry out such comparison in other ions with different degrees of ionization with 7--10 electrons.  

\bibliographystyle{aa_url}
{\footnotesize
\bibliography{references}}
\end{document}